\documentclass[
final
]{dmtcs-episciences}

\usepackage{epsfig }
\usepackage{latexsym}
\usepackage{amsfonts,amsmath,amssymb,textcomp}
\usepackage{algorithm}
\usepackage{algorithmic}
\allowdisplaybreaks

\newtheorem{thm}{Theorem}[section]
\newtheorem{lem}[thm]{Lemma}
\newtheorem{cor}[thm]{Corollary}
\newtheorem{prop}[thm]{Proposition}
\newtheorem{rem}[thm]{Remark}
\newtheorem{deff}[thm]{Definition}
\newtheorem{conj}[thm]{Conjecture}
\newtheorem{key}[thm]{Keywords}
\newtheorem{prob}[thm]{Problem}
\newenvironment{probb}{ \begin{prob} \rm}{ \end{prob} }
\newcommand{\bprobb}{\begin{probb}}
\newcommand{\eprobb}{\end{probb}}
\newcommand{\bth}{\begin{thm}}
\newcommand{\ethGL}{\end{thm}}
\newcommand{\bconj}{\begin{conj}}
\newcommand{\econj}{\end{conj}}
\newcommand{\bkey}{\begin{key}}
\newcommand{\ekey}{\end{key}}
\newcommand{\bl}{\begin{lem}}
\newcommand{\el}{\end{lem}}
\newcommand{\bdeff}{\begin{deff}}
\newcommand{\edeff}{\end{deff}}
\newcommand{\bcor}{\begin{cor}}
\newcommand{\ecor}{\end{cor}}
\newcommand{\bprop}{\begin{prop}}
\newcommand{\eprop}{\end{prop}}
\newcommand{\brem}{\begin{rem}}
\newcommand{\erem}{\end{rem}}
\newcommand{\beq}{\begin{equation}}
\newcommand{\eeq}{\end{equation}}
\newcommand{\beqn}{\begin{eqnarray}}
\newcommand{\eeqn}{\end{eqnarray}}
\newcommand{\beqns}{\begin{eqnarray*}}
\newcommand{\eeqns}{\end{eqnarray*}}
\newcommand{\ba}{\begin{array}}
\newcommand{\ea}{\end{array}}
\newcommand{\bit}{\begin{itemize}}
\newcommand{\eit}{\end{itemize}}
\newcommand{\bpr}{\noindent{\bf Proof.\hspace{1em}}}
\newcommand{\epr}{\hfill\rule{3mm}{3mm}\vspace{\baselineskip}\\}
\newcommand{\BO}{\mathcal{O}}

\newcommand{\non}{\nonumber}
\newcommand{\babs}{\begin{abstract}}
\newcommand{\eabs}{\end{abstract}}
\newcommand{\bal}{\begin{align}}
\newcommand{\bals}{\begin{align*}}
\newcommand{\bs}{\begin{skip}}
\newcommand{\eal}{\end{align}}
\newcommand{\eals}{\end{align*}}
\newcommand{\es}{\end{skip}}

\newcommand{\ra}{\rightarrow}

\newcommand{\B}{\quad}

\newcommand{\lp}{\left (}
\newcommand{\rp}{\right )}
\newcommand{\lb}{\left [}
\newcommand{\rb}{\right ]}

\def\ii{\mathbf{i}}

\newcommand{\mut}{\tilde{\mu}}
\newcommand{\sigt}{\tilde{\sig}}
\newcommand{\alt}{\tilde{\al}}

\newcommand{\pt}{\tilde{p}}

\newcommand{\mt}{\tilde{m}}

\newcommand{\nt}{\tilde{n}}

\newcommand{\wt}{\tilde{w}}

\newcommand{\Ln}{\lg n}

\newcommand{\chil}{\chi_l}

\newcommand{\De}{\Delta}
\newcommand{\K}{\kappa}

\newcommand{\al}{\alpha}

\newcommand{\Fi}{\varphi}
\newcommand{\FI}{\phi}

\newcommand{\II}{\infty}

\newcommand{\Tet}{\Theta}
\newcommand{\sig}{\sigma}
\newcommand{\sigd}{\sigma^2}

\newcommand{\gam}{\gamma}
\newcommand{\Gam}{\Gamma}

\newcommand{\ta}{\tau}

\def\1{{\ifmmode 1\mskip-1.5\thinmuskip\mathrm{l}%
        \else\textrm{1\hskip -.23em l}\fi}}

\newcommand{\E}{\mbox{$\mathbb E$}} 
\newcommand{\V}{\mbox{$\mathbb V$}} 
\def\P{{\mathbb {P}}}

\newcommand{\bin}[2]
{
{#1\choose #2}
}

\newcommand{\Std}[2]
{
\begin{Bmatrix} #1\\#2 
\end{Bmatrix}
}

\title{\bf The Adaptive Sampling Revisited }
\author{Matthew Drescher \affiliationmark{1}\thanks{supported  by ERC Consolidator Grant 615640-ForEFront, knavely@gmail.com}  \and Guy Louchard \affiliationmark{1} \thanks{louchard@ulb.ac.be} \and
Yvik Swan \affiliationmark{2} \thanks{yswan@ulg.ac.be}
}
\affiliation{
 Universit\'e Libre de Bruxelles,  Belgium\\
 Universit\'e de Li\`{e}ge, Belgium
  }
  
\date{\today}
\begin{document}
\received{2017-1-18}
\accepted{2019-4-18}
\publicationdetails{21}{2019}{3}{13}{2652}
\maketitle

\begin{abstract}
The problem of estimating the number
$n$
of
distinct
keys of a large collection
of
$N$
data is well known in computer science.   A  classical  algorithm  is the   adaptive
sampling   (AS). The cardinality  $n$
can  be  estimated  by $R\cdot 2^D$, where $R$
is  the final  bucket (cache) size and $D$ is the final depth at  the
end of the process. Several new interesting questions can be asked about AS (some of them were suggested by P.Flajolet and popularized by J.Lumbroso). The distribution of $W=\log (R2^D/n)$ is known,  we rederive this distribution in a simpler way. We provide new results on the moments of $D$ and $W$. We also analyze the final cache size $R$ distribution. We consider   colored keys:  assume  that among the $n$ distinct keys, $n_C$ do have color $C$.  We show how to estimate $p=\frac{n_C}{n}$. We also study colored keys with some  multiplicity given by some distribution function.  We want to estimate mean and variance of this distribution. Finally, we consider  the case where neither colors nor multiplicities are known. There  we want to estimate the related parameters.  An appendix is devoted to the case where the hashing function provides bits with probability different from $1/2$. 
\end{abstract}

\section{Introduction}\label{S1}

The problem of estimating the number
$n$
of
distinct
keys of a large collection
of
$N$
data is well known in computer science. It arises in query optimization of database  systems.  It has many practical applications. For example consider a stream (or log) of login events for a large popular website. We would like to know how many unique users login per month. A naive approach would be to insert each login credential into a set which is stored in memory. The cardinality of the set will of course be equal to the number of unique logins. However, if the number of distinct logins makes the cardinality of the set too large to fit into memory, this simple method will not work. While strategies involving more machines and/or writing to disk exist, see the paper by  Rajaraman and  Ullman \cite{RU2012}, the estimation technique we study here is an alternative requiring no additional infrastructure.

A classical algorithm is the adaptive sampling (AS), the original idea of which is due to Wegman in \cite{WE84}. The mean and variance
of AS are considered by   Flajolet  in \cite{FL90} .
Let  us  summarize the  principal  features  of AS. Elements  of the  given set  of
$N$
data  are  hashed  into  binary  keys.  These  keys  are  infinitely  long  bit  streams  such
that each bit has probability $1/2$
 of being $0$ or $1$.   A uniformity assumption is made
on the hashing function .

The algorithm keeps a  bucket (or cache)
$B$
of at most
$b$ distinct
keys. The depth of
sampling,
$d$
which  is  defined below, is  also  saved. We start  with
$d=0$ and throw
only
distinct keys
into
$B$. When
$B$
is  full,  depth
$d$
is  increased  by  $1$, the  bucket  is
scanned, and only keys starting with $0$ are kept.\\ (If the bucket is still full, we wait until a new key starting with $0$
appears. Then $d$ is again increased by $1$ and we keep only keys starting with $00$). The scanning on the set is resumed
and only distinct keys starting with $0$ are considered. More generally, at depth
$d$,
 only
distinct keys starting with $0^d$
are taken into account. When we have exhausted the
set  of
$N$
data,
$n$
can  be  estimated  by $R2^D$, where $R$ is the random final bucket (cache) size and $D$ is the final depth at the end of the process (total execution number of the process). 
 $R$ is the number of (all distinct) keys in the final cache and is immediately computed. We can summarize the algorithm with the following pseudo code

\medskip
\begin{algorithm}
\caption{}  %
\label{alg1}
\begin{algorithmic}
\medskip
\STATE {\bf Parameter}: bucket (or cache) $B$ of at most $b$ distinct keys.%
\STATE {\bf Input:} a stream $S=(s_1,s_2,\ldots,s_N)$ %
\STATE {\bf Output:} the final  bucket   size $R$  and  the final depth $D$%
\STATE {\bf Initialization:} $B:=\emptyset$ and $d:=0$%
 \FORALL{$x\in S$}
 \IF{$h(x)=0^d\ldots $}
\IF{$x\notin B$}
\STATE $B:=B\cup x$ %
\ENDIF;
\ENDIF;
\IF{$|B|>b \mbox{ (overflow of cache)    }$}  %
\STATE $d:=d+1$%
\STATE  filter $(B)\B$(remove keys  which  hash value doesn't match $0^d\ldots$)%
\ENDIF;
\ENDFOR;
\STATE $D:=d$%
\STATE \textbf {return} $R,D$;
\end{algorithmic}
\end{algorithm}

AS has some advantages in terms of processing time and conceptual simplicity.
     As  mentioned  in   \cite{FL90} (see also the paper by   Astrahan et al. \cite{AS87}),  AS  outperforms
standard sorting methods by a factor of about $8$. In terms of storage consumptions,
using $100$ words of memory will provide for a typical accuracy of $12\%$. This is to be 
contrasted  again  with  sorting,  where  the  auxiliary  memory  required  has  to  be  at
least as large as the file itself. Also an exact algorithm using a hash table will need a huge auxiliary memory.
Finally  AS  is  an  unbiased  estimator  of  cardinalities  of  large  files  that  necessitates
minimal auxiliary storage and processes data in a single pass.

In a paper by Gibbons \cite{GI01} we are introduced to the Distinct Sampling approach for distinct value queries and reports over streams with known error bounds. This approach is based on adaptive selection of a sample during one single pass through the data and is very similar conceptually to AS. This sample is then used to estimate key queries such as ``count distinct'' or {\em how many distinct values satisfy a given predicate?}, and ``Event Reports'' or  pre-scheduled, hard coded queries.

In fact \cite{GI01} shows experimental results which are more than 5 times more accurate than previous work and 2-4 orders of magnitude faster. This work is currently being considered for improved implementation of the widely used open source Postgres SQL database. See \cite{PSQL}. 

Several new interesting questions can be asked about AS (some of them were suggested by P. Flajolet and popularized by J. Lumbroso). The distribution of $W=\log (R2^D/n)$ is known (see \cite{GL97}), but in Sec.\ref{S3}, we rederive this distribution in a simpler way. In Sec.\ref{S4} we provide new results on the moments of $D$ and $W$. The final cache size $R$ distribution is analyzed in Sec.\ref{S5}. Colored keys are considered in Sec.\ref{S6}: assume that we have a set of colors and that each key has some color. 
Assume also that among the $n$ distinct keys, $n_C$ do have color $C$ and that $n_C$ is large such that $\frac{n_C}{n}=p=\Tet(1)$. We show how to estimate $p$. We consider keys with some  multiplicity in Sec.\ref{S7}:
assume that, to each key $\K_i$, we attach  a counter giving its \emph{observed}  multiplicity $\mu_i$. Also we assume that the multiplicities of  color $C$ keys are given by iid random variables (RV), with distribution function $F_C$,  mean $\mu_C$, variance $\sigd_C$ (functions of $C$). We show how to  estimate $\mu_C$ and  $\sigd_C$ .  Sec.\ref{S8} deals with the case where  neither colors nor multiplicities are known. We want to estimate color of keys, their multiplicities and their number.  An appendix is devoted to the case where the hashing function provides bits with probability different from $1/2$. 
\section{Preliminaries.}\label{S2}

Let us first give the main notations we will use throughout the paper. Other particular  notations will be provided where it is needed.
\bals
N&:= \mbox{ \emph{total} number of keys},\  N \mbox{ large}, \\ 
n&:=\mbox{number of \emph{distinct} keys},\ n\mbox{ large},  \\
\sim &:=\mbox{ asymptotic to, for large }n,\\
b&:=\mbox{ cache size},\ b\mbox{ fixed, independent of }n,  \\
\stackrel{b}{\sim} &:= \mbox{ asymptotic to, for large } n \mbox{ and }b,\\
R&:=\mbox{ number of keys in the cache, at the end of the process},\\
D&:= \mbox{ depth of the cache, at the end of the process},\\
Z&:=\frac{R 2^D}{n},\\
\lg &:=\log_2,\\
W&:=\lg (Z),\\
\end{align*} 
 
 Flajolet gives   the exact distribution in \cite{FL90}
\bal
p(r,d )&:=\P(R=r,D=d)= \bin nr \Big( \frac{1}{2^d} \Big)^r \Big(1-\frac{1}{2^d} \Big)^{n-r}
\bigg[1-\sum_{k=0}^{b-r}\bin {n-r}k \Big( \frac{1}{2^d} \Big)^k 
\Big(1-\frac{1}{2^d} \Big)^{n-r-k}\bigg],                                              \label{E00} \\ 
p(.,d)&:=\P(D=d)= \sum_{r=0}^b p(r,d),   \non\\
p(r,.)&:=\P(R=r)=\sum_d p(r,d), \non\\
P(r,d)&:=\P(R=r,D\leq d). \non
\end{align} 
The sample of $R$ elements at the end of the execution is random as the hashed keys are $i.i.d$ random variables: AS produces random samples. 

We can now see Adaptive Sampling as an urn model, where balls (keys), are thrown into urn $D=d$ with probability $1/2^d$. We recall the main properties of such a model.

\bit
\item 
\textsc{Asymptotic independence.} We have asymptotic independence of urns, for all events related to urn $d$  ($d$ large) containing $\BO(1)$ balls. 
This is proved, by Poissonization-De-Poissonization, in \cite{HL00},   \cite{ALP05} and  \cite{LP05}. This technique can be biefly described as follows. First we construct the corresponding generating function. Then we Poissonize (see, for instance, the paper by Jacquet and Szpankowski \cite{JS98} for a general survey): instead of using a \emph{fixed}  number of balls, we use $N$ balls, where $N$ is a Poisson random variable. It follows that the urns become \emph{independent} and the number of balls in urn $\ell$ is  a Poisson random variable. We turn then to complex variables, and with Cauchy's integral theorem, we De-Poissonize the generating function, using 
\cite[Thm.10.3 and Cor.10.17]{JS98}. The error term is $\BO(n^{-\gam})$ where $\gam$ is a positive constant.
\item \textsc{Asymptotic distributions.} We obtain asymptotic distributions of the interesting random variables  as follows. The number of balls in each urn is asymptotically Poisson-distributed with parameter $n/2^d$ in  urn $d$ containing $\BO(1)$ balls (this is the classical asymptotic for the Binomial distribution). This means that the asymptotic number $\ell$ of balls in urn $d$ is given by 
\[\exp \lp -n/2^d \rp\frac{\lp n/2^d\rp^\ell}{\ell !},\]
and with $\eta=d-\lg n,L:=\ln 2$, this is equivalent to a Poisson distribution with parameter $ e^{-L\eta}$. The asymptotic distributions are related to Gumbel distribution functions (given by $\exp\lp -e^{-x}\rp$) or convergent series of such. The error term is  $\BO(n^{-1})$. 
\item \textsc{Uniform Integrability.} We have uniform integrability for the moments of our random variables. To show that the limiting moments are equivalent to the moments of the limiting 
distributions, we need a suitable rate of convergence. This is related to a uniform integrability condition
(see Lo\`{e}ve's book  \cite[Section~11.4]{LO63}). For Adaptive Sampling, the  rate of convergence is analyzed in detail in \cite{momP}.
  The error term is  $\BO(n^{-\gam})$. 
 \item \textsc{Mellin transform.} Asymptotic expressions for the moments are obtained by Mellin transforms (for a good reference to 
Mellin transforms, see the paper by Flajolet et al. \cite{FGD95}). The error term is  $\BO(n^{-\gam})$.  We proceed as follows (see \cite{momP} for detailed proofs): from the asymptotic properties of the urns, we have obtained  the asymptotic distributions of our random variables of interest. Next we compute the Laplace transform $\FI(\al)$ of these distributions, from which we can derive the dominant part of probabilities and moments as well as the (tiny) periodic part in the form of a Fourier series. This connection will be detailed in the next sections. 
 \item \textsc{Fast decrease property.} The gamma function $\Gam(s)$ decreases exponentially in the direction  $i\II$:
\[|\Gam( \sig+\ii t )| \sim \sqrt{2\pi}|t|^{\sig-1/2}e^{-\pi |t|/2}.\]
Also, this property is true for all other functions we encounter. So inverting the Mellin transforms is easily justified.
\item \textsc{Early approximations.} If we compare the approach in this paper with other ones that appeared previously, then
we can notice the following. Traditionally, one would stay with exact enumerations as long
as possible, and only at a late stage move to asymptotics. Doing this, one would, in terms
of asymptotics, carry many unimportant contributions around, which makes the computations
quite heavy, especially when it comes to higher moments. Here, however, approximations are
carried out as early as possible, and this allows for streamlined (and often automatic)
computations of the higher moments. 
 \eit

We set  $\eta=d-\Ln$, (\ref{E00})  leads to
\beq
 p(r,d) \sim f(r,\eta)= \exp(- 2^{-\eta} ) \frac{ 2^{-r\eta}}{r!} 
\bigg[1-\exp(- 2^{-\eta} ) \sum_{k=0}^{b-r}  \frac{ 2^{-k\eta}}{k!} \bigg] ,                \label{E0}
\eeq
and similar functions for $P(r,d )$. Asymptotically, the distribution will be a periodic function of the fractional part of $\lg n$.
The distribution $P(r,d )$ does not converge in the weak sense, it does however converge  
 along subsequences $n_m$ for which the fractional part of $\lg n_m$ is constant. 
 This type of convergence is not uncommon in the Analysis of Algorithms. Many examples are given in \cite{momP}.
 
 From (\ref{E0}),  we compute the Laplace transform, with $\alt :=\al/L$:
 \[ \FI(r,\al) =\int_{-\II}^\II e^{\al \eta}f(r,\eta)d\eta=
 \frac{\Gam(r-\alt)}{Lr!}
 -\sum_{k=0}^{b-r}\frac{\Gam( r+k-\alt )}{Lr!k!2^{r+k-\alt }} . \]
 The $k$-th  moments of $Z$ are already given in \cite{GL97} and \cite{momP}. As shown in \cite{GL97}, we must have $k\leq b$. For the sake of completeness, we repeat them here, with $\chil :=\frac{2l\pi \ii}{L}$,  $\Std{k}{i}$  denoting the Stirling number of the second kind, and $\V(X)$ denoting the  Variance of  the random variable $ X$: 
 \bals
 \E[Z^k]&\sim  m_{1,k}+w_{1,k},\\
 m_{1,k}&=1+\frac{(b-k)!}{L}\sum_{i=1}^{k-1}\Std{k}{i}\frac{2^{k-i}-1}{(k-i)(b-i)!},\\
 w_{1,k}&=\sum_{l\neq 0}  \frac1L\sum_{j=1}^{k-1}\Std{k}{j}
\lb(1-2^{k-j}\rb
\Gamma(j-k+\chil)\binom{b-k+\chil}{b-j}e^{-2l\pi\ii \lg n},
\end{align*} 
\beq
m_{1,1}=1,w_{1,1}=0,m_{1,2}=1+\frac{1}{(b-1)L},\V(Z)\sim \frac{1}{(b-1)L}.      \label{E21}
\eeq
$w_{i,j}$ will always denote  a periodic function of  $\lg n$ of small amplitude.
 Note that, in   \cite{FL90}, Flajolet already computed $m_{1,1},m_{1,2},w_{1,1}\mbox { and }w_{1,2}$. 
\section{Asymptotic distribution of $W=D-\lg n +\lg R$}\label{S3}
$W$ corresponds to the bit size  of $Z$ and has some independent interest. 
Let us recover this distribution from (\ref{E0}). In the sequel, we will denote $\E(Y|A)P(A)$  by $\E(Y;A)$, with $Y$ either a Boolean event or a random variable.  
 We have the following theorem
 
\bth
The asymptotic distribution of $W=D-\lg n +\lg R$, with $R>0$ is given by
 \[\P(W\leq \al,R>0)\sim \sum_{r=1}^b \sum_{l\geq 0} \exp(- 2^{-\Fi} ) \frac{ 2^{-r\Fi}}{r!} 
\lb 1-\exp(- 2^{-\Fi} ) \sum_{k=0}^{b-r}  \frac{ 2^{-k\Fi}}{k!}\rb,\]
where
\[\{x\} :=\mbox{ fractional part of }x,\]
\[\Fi:=\lfloor \{\lg n\}-\lg r+\al  \rfloor-\{\lg n\}-\ell.\]
\ethGL
\bpr
 \bals
 \P(W\leq \al,R>0)&=\P[D\leq \lg n-\lg R+\al,R>0]\\
 &=\P[D \leq \lfloor \lg n \rfloor +\lfloor \{\lg n\}-\lg R+\al  \rfloor,R>0]\\
 &\sim \sum_{r=1}^b \sum_{\ell\geq 0} \exp(- 2^{-(\eta-\ell)} ) \frac{ 2^{-r(\eta-\ell)}}{r!} 
\lb 1-\exp(- 2^{-(\eta-\ell)} ) \sum_{k=0}^{b-r}  \frac{ 2^{-k(\eta-\ell)}}{k!}\rb,
 \end{align*} 
 with
 \[\eta=\lfloor \{\lg n\}-\lg r+\al  \rfloor-\{\lg n\},\]
 or
 \[\P(W\leq \al,R>0)\sim \sum_{r=1}^b \sum_{l\geq 0} \exp(- 2^{-\Fi} ) \frac{ 2^{-r\Fi}}{r!} 
\lb 1-\exp(- 2^{-\Fi} ) \sum_{k=0}^{b-r}  \frac{ 2^{-k\Fi}}{k!}\rb,\]
with
\[\Fi:=\lfloor \{\lg n\}-\lg r+\al  \rfloor-\{\lg n\}-\ell.\]
This is exactly Theorem   4.1 in \cite{GL97} that we obtain here in a simpler way.
\epr

\section{Moments of $D-\lg n$ and $W$}\label{S4}
Recall that $D$ is the final depth (number of times the cache is filtered = number of times the cache overflows).
Two interesting parameters are given by the moments of $D-\lg n$ and $W$. Their asymptotic behaviour is given as follows, with 
$\psi(x)$ denoting the digamma function (the logarithmic derivative of $\Gam(x)$)
\bth
 The moments of $D-\lg n$ and $W$ are asymptotically given by
\[\E[(D-\lg n)^k;R=r]\sim \mt_{k,r}+\wt_{k,r},\] 
 where
\bals
\mt_{k,r}&:=\FI^{(k)}(r,0),\\
w_{k,r}& =\sum_{l\neq 0}  
 \left. \FI^{(k)}(r ,\al)\right|_{\al=-L \chil }  e^{-2l\pi\ii \lg n}.\\
&\mbox{For instance}\\
 \mt_{1,r}&=-\frac{\psi(r)}{L^2 r}+\sum_{k=0}^{b-r}\frac{(\psi(r+k)-L)2^{-(r+k)}\Gam(r+k)}{L^2\Gam(r+1)\Gam(k+1)},\B r>0,\\
 \mt_{1,0}&=\frac12+\frac{\gam}{L}+\sum_{k=1}^{b}\frac{(\psi(k)-L)2^{-k}}{kL^2},\\
 \tilde{w}_{1,r} &=\sum_{l\neq 0} \lb -\frac{\psi(r+\chil)\Gam(r+\chil)}{L^2\Gam(r+1)}+ \sum_{k=0}^{b-r}
  \frac{(\psi(r+k+\chil)-L)2^{-(r+k)}\Gam(r+k+\chil)}{L^2\Gam(r+1)\Gam(k+1)} \rb e^{-2l\pi\ii \lg n},\B r>0,\\
   \tilde{w}_{1,0}&=\sum_{l\neq 0}\lb  -\frac{\psi(\chil)\Gam(\chil)}{L^2}+ \sum_{k=0}^{b} \frac{\Gam(k+\chil)}{L^2k!2^k}(\psi(k+\chil)-L)\rb.
 \end{align*} 
 
 \bals
  \E(W;R>0)&\sim \sum_{r=1}^b \mt_{1,r}+\sum_{r=1}^b p(r,.)\lg r +\sum_{r=1}^b \wt_{1,r},\\
 \E(W^2;R>0)&\sim \sum_{r=1}^b  \mt_{2,r} +2 \sum_{r=1}^b \mt_{1,r}\lg r +  \sum_{r=1}^b p(r,.)(\lg r)^2
 +\sum_{r=1}^b \wt_{2,r}+   2\sum_{r=1}^b \wt_{1,r} \lg r.\\
  \end{align*} 
\ethGL
\bpr
 Using the techniques developed   in \cite{momP}, we obtain the dominant (constant) part of the moments of $D$ as follows:
 \[\E[(D-\lg n)^k;R=r]\sim \mt_{k,r}+w_{k,r},\]
 where the non-periodic component is given by 
\[\mt_{k,r}:=\FI^{(k)}(r,0),\]
 and the corresponding periodic term  is given by
 \[w_{k,r} =\sum_{l\neq 0}  
 \left. \FI^{(k)}(r ,\al)\right|_{\al=-L \chil }  e^{-2l\pi\ii \lg n}.\]
 This was already computed in \cite{momP}, but with some errors. The first corrected values are now provided.
 
 As  $W=D-\lg n+\lg R$, the rest of the proof  is immediate
\epr
 
 It will be useful to obtain an asymptotic for the expectation of $D-\lg n$ (non-periodic component) for large $b$. This is computed as follows. First of all, we rewrite 
 $\sum_{r=1}^b \mt_{1,r}$ as
 \[\sum_{r=1}^b \mt_{1,r} =-\sum_{r=1}^b \frac{\psi(r)}{L^2 r}
 + \sum_{u=1}^b \lb \sum_{r=1}^u \frac{1}{\Gam(r+1)\Gam(u-r+1)}\rb\frac{(\psi(u)-L)2^{-u}\Gam(u)}{L^2}.\]
 Now it is clear that the main contribution of the second term is related to large $u$. So we set $r=\frac{u}{2}+v$. This gives, by Stirling,
 \[\Gam(r+1)\sim e^{-(u/2+v)}e^{v+v^2/u}\lp \frac{u}{2}\rp^{u/2+v}\sqrt{\pi u},\]
 and
 \[\Gam(r+1)\Gam(u-r+1)\sim e^{-u}e^{2 v^2/u}\lp \frac{u}{2}\rp^{u}\pi u.\]
 By  Euler-Maclaurin, we have
\[\sum_{r=1}^u \frac{1}{\Gam(r+1)\Gam(u-r+1)}\sim 2\sum_{v=0}^{u/2} \frac{e^u }{\lp \frac{u}{2}\rp^{u}\pi u}e^{-2 v^2/u}
\sim\int_0^\II 2 e^{-2 v^2/u} dv \frac{e^u }{\lp \frac{u}{2}\rp^{u}\pi u}=\frac{e^u }{\lp \frac{u}{2}\rp^{u}\sqrt{2\pi u}},\] 
and, finally,
\[\sum_{r=0}^b \mt_{1,r}\stackrel{b}{\sim} \frac12+\frac{\gam}{L}+\sum_{u=1}^b \lb -\frac{\psi(u)}{L^2 u} +\frac{(\psi(u)-L)2^{-u}}{uL^2}+ 
\frac{(\psi(u)-L)2^{-u}\Gam(u)}{L^2}\frac{e^u }{\lp \frac{u}{2}\rp^u\sqrt{2\pi u}}\rb,\]
and, to first order,
\beq
\E(D-\lg n)\sim\sum_{r=0}^b \mt_{1,r}\stackrel{b}{\sim} \frac12+\frac{\gam}{L}-\sum_{u=1}^b\frac{1}{Lu}\sim -\lg b+\BO(1)               \label{E10} 
\eeq
The expected  total time cost of the algorithm, $\mathbf{C}_n$, is given by 
\[\E[\mathbf{C}_n]= n \BO(\lg b)+\E[D] \BO(b) = n\BO(\lg b)+ \lg n  \BO(b),\]
where $\BO(\lg b)$ is the update cost of the cache for each key (we assume an efficient implementation of the cache, for instance a binary search  tree) and $\BO(b)$ is the update cost of the cache at each process execution.
\section{Distribution of $R$}\label{S5}
The asymptotic moments and distribution of the cache size $R$ are given as follows
\bth
The non-periodic components of the asymptotic moments and distribution of $R$ are given by
\bals
\E(R)&\sim
\frac{b}{2L},\\
\E(R^2)&\sim \frac{b(3b+1)}{8L},\\
\V(R)&\sim \frac{b(3Lb-2b+L)}{8L^2},\\
\P(R=r)&= p(r,.)\sim \frac1L\lb \frac1r- \sum_{k=0}^{b-r}\frac{\Gam(r+k)}{r!k!2^{r+k}} \rb,r\geq 1,
\end{align*} 
Similarly, the periodic components are given by
\bals
w_1(R)&=\sum_{r=0}^b r \sum_{l\neq 0} \FI(r,-L\chil)   e^{-2l\pi\ii \lg n},\\
w_2(R)&=\sum_{r=0}^b r^2 \sum_{l\neq 0} \FI(r,-L\chil)   e^{-2l\pi\ii \lg n},\\
w_0(r)&=  \sum_{l\neq 0} \FI(r,-L\chil)   e^{-2l\pi\ii \lg n}
\end{align*} 

\begin{figure}[htbp]
	\centering
		\includegraphics[width=0.8\textwidth,angle=0]{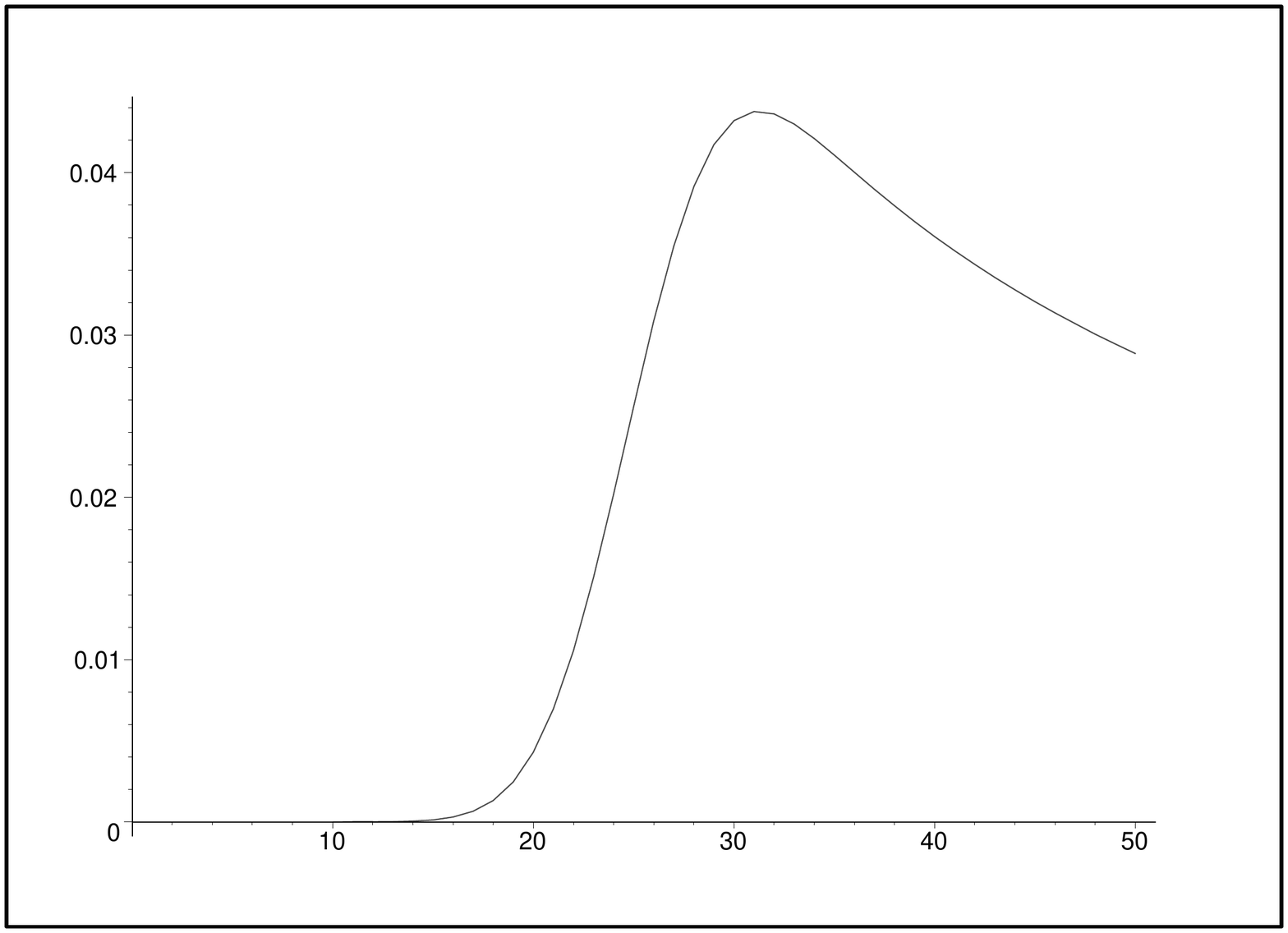}
	\caption{  $p(r,.)$ for $b=50$}
	\label{F1}
\end{figure}

\ethGL
\bpr
We have
\[\P(R=r)= p(r,.)\sim  \FI(r,0)=\frac1L\lb \frac1r- \sum_{k=0}^{b-r}\frac{\Gam(r+k)}{r!k!2^{r+k}} \rb,r\geq 1,\]
and $p(0,.)=1-\sum_1^b p(r,.)$, with
\bals
\sum_1^b p(r,.)&\sim \frac1L\lb  H_b- \sum_{r=1}^b \sum_{k=0}^{b-r}\frac{\Gam(r+k)}{r!k!2^{r+k}} \rb \\
&=\frac1L\lb  H_b- \sum_{u=1}^b  \frac{(u-1)!}{2^u}   \sum_{r=1}^u    \frac{1}{r!(u-r)!}                     \rb,
\end{align*} 
Where $H_n$ denotes the $n-th$ harmonic number.
This quantity was already obtained in \cite{GL97} after some complicated algebra! This leads to
\[p(0,.)\sim 1-\sum_{u=1}^b\frac{1}{u 2^uL},\]
which is also the probability of $Z=0$.
This is also easily obtained from $\lim_{r\ra 0}\FI(r,0)$. Figure \ref{F1} gives $p(r,.)$ for $b=50$

The moments of $R$ are computed as follows.
\bals
\E(R)&= \sum_{r=1}^b r p(r,.)\sim \frac1L\lb  b- \sum_{u=1}^b  \frac{(u-1)!}{2^u}   \sum_{r=1}^u    \frac{r}{r!(u-r)!}                \rb\\
&=\frac1L\lb  b- \sum_{u=1}^b  \frac{1}{ 2^u}   [2^{u-1}]                \rb\\
&=\frac{b}{2L}.
\end{align*} 
More generally, the generating function of $p(r,.)$ is given by
\bals
\sum_{r=1}^b z^r p(r,.)&\sim \frac1L\lb  \sum_{r=1}^b\frac{z^r}{r}- \sum_{u=1}^b  \frac{(u-1)!}{2^u}   \sum_{r=1}^u    \frac{z^r}{r!(u-r)!}     \rb\\
&=\frac1L\lb  \sum_{r=1}^b\frac{z^r}{r}- \sum_{u=1}^b  \frac{1}{u 2^u}[(1+z)^u-1]                                       \rb.
\end{align*} 
This leads to
\bals
\E(R^2)&\sim \sum_{r=1}^b r^2 p(r,.)=\frac{b(3b+1)}{8L},\\
\V(R)&\sim \frac{b(3Lb-2b+L)}{8L^2}.
\end{align*} 
Similarly, the periodic components are given by
\bals
w_1(R)&=\sum_{r=0}^b r \sum_{l\neq 0} \FI(r,-L\chil)   e^{-2l\pi\ii \lg n},\\
w_2(R)&=\sum_{r=0}^b r^2 \sum_{l\neq 0} \FI(r,-L\chil)   e^{-2l\pi\ii \lg n}.
\end{align*} 
\epr
\section{Colors}       \label{S6}
Seasonal, or temporal context, has become increasingly important in data mining \cite{season}. For example its often important to be able to group events by the time which they occur or to understand event periodicity. We can represent different temporal contexts with colors.  This motivates our analysis of colored keys.

Assume that we have a set of colors and that each key has some color. Let us give a simple example. We might be interested in knowing, for instance, the proportion  $p$ of elements  whose multiplicity  is below some constant $M$. So we say that such elements are of color  ``white'', while the rest are of color ``black''.  We attach to each \emph{distinct} key 
$\K_\ell$ a counter $\nu_\ell$ giving the number of times (multiplicity)  this key appears in the sample. At the end of AS, we have in the cache $U_W$ white keys. AS produces random samples and it can gather exact counts of the frequency of the sampled elements (since a sampled element enters the sample in its very first occurrence  and if an element is kicked out from the sample, it will never be sampled again).

This leads to the unbiased estimates $\nt_W=2^D U_W,\nt=2^D R,\pt_W=U_W/R$.

We note that the observed multiplicities will be used with more detail in Sec. \ref{S7}, \ref{S8}.

Another example is as follows.
A situation naturally maps to real life situations when data is ``strongly seasonal''. Such is the case when noisy ``seasonal'' differences in data are to be naturally expected, and therefore to be ignored when performing analytics of a data stream. For example, even though the number of viewers on two consecutive baseball games might show the second game receiving roughly half as many views as the first, it makes little sense to conclude that the baseball team is becoming less popular if the first was a night game and the second a weekend day game. See the report by Wong et al. \cite{NFLX} for an example of outlier detection where seasonality must be ignored.

Assume now that among the $n$ distinct keys, $n_C$ do have color $C$ and that $n_C$ is large such that $\frac{n_C}{n}=p=\Tet(1),q:=1-p$.  In the cache, the $R$ keys (we assume $R>0$) contain $U$ keys with color $C$ with probability distribution
\[\P(U=u|R=r)=\frac{\bin{n_C}u\bin{n-n_C}{r-u}}{\bin nr}, \]
and, if $r=o(n)$, this is asymptotically given by the conditioned binomial distribution $Bin(r,p)$. We want to estimate $p$. We are interested in the  distribution of the statistic $\pt=U/R$ . We have 
\bth
The asymptotic moments of the statistic $\pt=U/R$ are given by 
\bal
\E\lp \frac{U}{R};R>0\rp &\sim p,\non\\
\E\lp \lp \frac{U}{R}\rp^2;R>0\rp &\sim p^2+pq\E\lp \frac1R;R>0\rp       ,   \label{E2} \\
\V\lp \frac{U}{R};R>0\rp &\sim   pq\E\lp \frac1R;R>0\rp  \non .                              
\end{align} 
\ethGL
\bpr
We have
\beq
\P\lb \lp \frac{U}{R}\rp\leq \al;R>0\rb =\P(U\leq \al R;R>0] \sim \sum_{r=1}^b p(r,.)\sum_{u=0}^{\lfloor \al r \rfloor} \bin ru p^uq^{r-u}  .              \label{E1}
\eeq
Now, conditioned on $R$, we have
\[\E(U|R)\sim Rp,\E(U^2|R)\sim Rpq+R^2p^2.\]
So, conditioned on $R$,
\bals
\E\lp \frac{U}{R}\rp&\sim p,\\
\E\lp \lp \frac{U}{R}\rp^2\rp &\sim p^2+\frac{pq}{R},
\end{align*} 
and, unconditioning leads to the theorem.
\epr

Intuitively, if the cache size $b$ is large, we should have an asymptotic Gaussian distribution for $U/R$. Actually, the fit is quite good, even for $b=30$ (and $p=0.2$).

This is proved in the next subsections.

\subsection{The distribution of $U/R$ for large $b$
 .}       \label{S61}

Let $R$  be a (possibly degenerate) random variable taking values
on  the (strict) positive integers. Conditionning on $R$, let  $U \sim Bin(R,p)$ for some known
$0<p<1$, and set $Y = U/R$. It appears that, as $R$ grows
large, the distribution of $Y$ becomes
asymptotically Gaussian. This claim can be made precise as follows. 
\bth   \label{T1}
Let $V \sim \mathcal{N}(0, 1)$ and write $ Y^* = \sqrt R \frac{Y-p}{\sqrt{pq}}.$
Then there exists an
absolute constant $\kappa \in \mathbb{R}$ such that 

\[d_{\mathcal{W}} \left( Y^*, V  \right)  \le \kappa \E \left\{ \frac{1}{\sqrt R} \right\}\]

for $d_{\mathcal{W}}\left( Y^*, V  \right)$ the Wasserstein distance
between the law of  $Y^*$ and that of $V$; moreover this constant is
such that 
 \[ \kappa \le \frac{q^2- p^2}{ \sqrt{pq}}
  + 4 \lb \frac{p^3+q^3}{pq}-1\rb^{1/2}. \]

\ethGL

\




\bpr
We will prove this theorem using the Stein methodology which, for $h
\in \mathcal{H}$, ($\mathcal{H}$ is a nice class of test functions),  suggests to write 
\begin{align*}
  \E h(Y^*) - \E h(V) &= \E \left( Y^* f(Y^*) - f'(Y^*) \right)
\end{align*}
 with $f := f_h$ such that 
 \begin{equation}
   \label{eq:4}
xf(x) -  f'(x) = h(x) - \E h(V).
 \end{equation}
(this is known as the \emph{Stein equation} for 
the Gaussian distribution)  so that
\begin{align}\label{eq:2}
  d_{\mathcal{W}}(Y^*, V) & = \sup_{h \in \mathcal{H}} \E \left|
    h(Y^*) - h(V) \right| \le \sup_{f_h} \left| \E \left( Y^* f(Y^*)
      - f'(Y^*) \right) \right|.  
\end{align}
  The reason why \eqref{eq:2}  is interesting is that    properties of
  the solutions  $f$ of \eqref{eq:4} are well-known  -- see, e.g.,
  \cite[Lemma 2.3]{BC105}  and  \cite[Lemma 2.3]{BC205} --  and quite
  good so that they can be used with quite some efficiency to tackle
  the rhs of  \eqref{eq:2}.   In the present
configuration  we know that $f_h$ is continuous and bounded on
$\mathbb R$, with  
\begin{align}
\|f_h'\| & \le \min \left( 2 \|h-\E h(V)\|, 4 \|h'\| \right)  \label{eq:12}
\end{align}
and 
\begin{equation}
  \label{eq:19}
  \|f_h''\| \le 2 \| h'\|.
\end{equation}
 In particular, if $H$ is the class of Lipschitz-1 functions with $\|h'\| \leq 1$ (this class generates the Wasserstein distance) then
\[\|f_h'\|\leq 4 \mbox{ and }\|f_h''\|\leq 2. \]
These will suffice to our purpose.

Our proof follows closely the
standard one for independent summands (see, e.g., \cite[Section
3]{RO11}). First we remark that, given $R \ge 1$, we can write  $Y^*$
as 
\begin{equation*}
  Y^* = \frac{1}{\sqrt R}\sum_{i=1}^R \xi_i
\end{equation*}
where, taking $X_i$ i.i.d. $Bin(1, p)$, we let 
$\xi_i=
({X_i-p})/{\sqrt{pq}}$ (which are centered and have
variance 1).  Next, for $r\ge 1$ and $1 \le i \le r$, define
\begin{align*}
  Y_i^{*r} =  Y^*-\frac{1}{\sqrt{r}}\xi_i = \frac{1}{\sqrt r}
  \sum_{j\neq i} \xi_{j}. 
\end{align*}
Next take $f$ solution of  (\ref{eq:4}) with $h$ some  Lipschitz-1 function. Then note that
 $ \E \left\{  \xi_i f \left( Y_i^{*r} \right)
  \right\} =0$
for all $1 \le i \le r$. 
We abuse
notations and, given $R$, write $Y_i^{*R} = Y^*_i$. Then
\begin{align*}
  \E \left\{ Y^* f (Y^*) \, | \, R\right\} & = \E \left\{\left.
    \frac{1}{\sqrt R} \sum_{i=1}^R \xi_if(Y^*)  \, \right| \, R
\right\}\\
& =  \E \left\{ \left. \frac{1}{\sqrt R} \sum_{i=1}^R \xi_i
  \left( f(Y^*) - f(Y^*_i) \right)  \, \right| \, R \right\} \\
& =  \E \left\{ \left. \frac{1}{\sqrt R} \sum_{i=1}^R  \xi_i
  \left( f(Y^*) - f(Y^*_i) - (Y^*-Y^*_i)f'(Y^*) \right)  \,
  \right| \, R \right\}   \\
& \quad \quad \quad +  \E \left\{ \left. \frac{1}{\sqrt R}
    \sum_{i=1}^R   \xi_i
  (Y^*-Y^*_i)f'(Y^*)   \,  \right| \, R \right\}
\end{align*}
so that 
\begin{align*}
 | \E \left\{  Y^* f(Y^*) - f'(Y^*) \, | \, R \right\}| & \le
\E \left\{ \left. \frac{1}{\sqrt R} \sum_{i=1}^R \left|  \xi_i 
  \left(  f(Y^*) - f(Y^*_i) - (Y^*-Y^*_i)f'(Y^*) \right)  \right| \,
  \right| \, R \right\}  \\
& \quad \quad \quad +  \left| \E \left\{ \left. f'(Y^*) \left( 1-  \frac{1}{\sqrt
      R} \sum_{i=1}^R  \xi_i 
  (Y^*-Y^*_i) \right)  \,  \right| \, R \right\} \right|\\
& =: | \chi_1(R)|+| \chi_2(R)|.
\end{align*}
Recall that $Y^*-Y^*_i=
\frac{1}{\sqrt R} \xi_i$. Then  (by Taylor expansion) we can easily
deal with  the first term to obtain 
\begin{align*}
| \chi_1(R)| 
& = \frac{\|f''\|}{2}   \frac{1}{\sqrt R}  \E 
    \left| \xi_1\right|^3 .  
\end{align*}
Taking expectations with respect to $R$ and using \eqref{eq:19} we
conclude 
\begin{align}\label{eq:5}
  E  |\chi_1(R)| & \le  \E \left| \xi_1\right|^3  E \left( \frac{1}{\sqrt R} \right).
\end{align}
For the second term note how 
\begin{align*}
  |\chi_2(R)| & = \left|  \E \left\{  f'(Y^*) \left( 1- \frac{1}{R}
      \sum_{i=1}^R \xi_i^2 \right) \, | \, R \right\} \right|\\
& = \left| \E \left\{ \left.  \frac{f'(Y^*)}{R}  \sum_{i=1}^R \left( 1-
    \xi_i^2  \right) \, \right| \, R\right\}
\right|\\
& \le \frac{\| f'\|}{R}  \E \left\{ \left.  \left| \sum_{i=1}^R \left( 1-
     \xi_i^2  \right) \right|\,  \right| \,R
\right\}.
\end{align*}
Since $\E \left\{ \sum_{i=1}^R \left( 1-
     \xi_i^2  \right) \, | \, R \right\} = 0$ we can pursue to obtain 
\begin{align*}
  |\chi_2(R)| & \le \frac{\| f'\|}{R}  \sqrt{\V \left(
      \left. \sum_{i=1}^R \left( 1- 
     \xi_i^2  \right) \,  \right| \,R \right) }  \\
& = \frac{\| f'\|}{\sqrt R}  \sqrt{\V \left(  \xi_1^2 \right) }
\end{align*}
where we used (conditional) independence of the $\xi_i$. Taking expectations with respect to $R$ and using
\eqref{eq:12} we deduce (recall $\V \left(  \xi_1^2 \right)
= \E \xi_1^4 -1$) 
\begin{align}\label{eq:3}
\E|\chi_2(R)| & \le 4 \sqrt{\E \xi_1^4-1} \E \left( \frac{1}{\sqrt R} \right).
\end{align}
Combining \eqref{eq:5} and \eqref{eq:3} we can conclude
\begin{align*}
   d_{\mathcal{W}}(Y^*, V) \le \left( \E |\xi_1^3| + 4 \sqrt{\E \xi_1^4-1}
  \right) \E \left( \frac{1}{\sqrt R} \right). 
\end{align*}
The claim follows. 
\epr
So we need the moments of $1/R$ for large $b$ (we limit ourselves to the dominant term).

\subsection{Moments of $1/R,R>0$ for large $b$}       \label{S62}
We have the following property

\bth
The asymptotic moments of $1/R,R>0$ for large $b$, with $R>0$ are given by 
\[\E \lp \frac{1}{R^{\al}} ;R>0\rp  \stackrel{b}{\sim} \frac{1}{L\al b^\al}(2^\al-1),\al>0.\]
\ethGL
\bpr
We have
\bals
\E\lp \frac1R ;R>0\rp&= \sum_{r=1}^b  p(r,.)/r\sim \frac1L\lb  \sum_{r=1}^b\frac{1}{r^2} - \sum_{u=1}^b  \frac{(u-1)!}{2^u}   \sum_{r=1}^u    \frac{1}{rr!(u-r)!}                \rb\\
&=\frac1L\lb    H_b^{(2)}  - \sum_{u=1}^b  \frac{1}{2^u}      {}_2 F_2[[1,-u+1];[2,2];-1]                                    \rb\\
&=\frac1L\lb   -\psi(1,b+1)+\frac{\pi^2}{6}   - \sum_{u=1}^\II  \frac{1}{2^u}      {}_2F_2[[1,-u+1],[2,2],-1] \rb \\ 
&+ \frac1L\sum_{u=b+1}^\II  \frac{1}{2^u}      {}_2F_2[[1,-u+1],[2,2],-1].                                                               
\end{align*}
where $\psi(n,x)$ is  the $n$th polygamma function, that  is the $n$th derivative of the digamma function 
$\psi(x)=\Gam'(x)/\Gam(x)$  and 
${}_2F_2$ is  the hypergeometric function.

But\footnote{We are indebted to H.Prodinger for this identity}
\bal
&\sum_{u=1}^\II  \frac{1}{2^u}      {}_2F_2[[1,-u+1],[2,2],-1]  \non  \\
&=\sum_{r=1}^\II \sum_{u=r}^\II \frac{(u-1)!}{2^u rr!(u-r)!} \non\\
&=\sum_{r=1}^\II\frac{1}{r^2} \sum_{u=r}^\II \frac{(u-1)!}{2^u (r-1)!(u-r)!}\non \\
&=\sum_{v=0}^\II\frac{1}{(v+1)^2} \sum_{u=v+1}^\II \frac{(u-1)!}{2^u v!(u-v-1)!}\non \\
&=\sum_{v=0}^\II\frac{1}{(v+1)^2} \sum_{w=v}^\II \frac{w!}{2^{w+1} v!(w-v)!}\non \\
&=\frac12 \sum_{v=0}^\II\frac{1}{(v+1)^2} \sum_{w=v}^\II \bin wv 2^{-w}\non\\
&=\frac12 \sum_{v=0}^\II\frac{1}{(v+1)^2} \sum_{s=0}^\II \bin {s+v}v  2^{-(s+v)}\non\\
&=\frac12 \sum_{v=0}^\II\frac{1}{(v+1)^2}2^{-v} \sum_{s=0}^\II \bin {-v-1}s  (-2)^{-s}\non \\
&=\frac12 \sum_{v=0}^\II\frac{1}{(v+1)^2} 2^{-v}\lp1-\frac12 \rp^{-(v+1)}\non\\
&=\sum_{v=0}^\II\frac{1}{(v+1)^2}\non\\
&=\zeta(2)=\frac{\pi^2}{6} .                                                        \label{E5}
\end{align}
Now
\[\psi(1,b+1)\stackrel{b}{\sim} \frac1b+\BO\lp \frac{1}{b^2}\rp,\]
and
\bals
&\sum_{u=b+1}^\II  \frac{1}{2^u}      {}_2F_2[[1,-u+1],[2,2],-1] =T_1+T_2, \\
&\mbox{where}\\
T_1&=\sum_{r=1}^{b+1} \sum_{u=b+1}^\II \frac{(u-1)!}{2^u rr!(u-r)!} \\
&=\frac12 \sum_{v=0}^b\frac{1}{(v+1)^2} \sum_{w=b}^\II \bin wv 2^{-w},\\
T_2&=\sum_{r=b+1}^\II  \sum_{u=r}^\II \frac{(u-1)!}{2^u rr!(u-r)!} \\
&=\frac12 \sum_{v=b}^\II\frac{1}{(v+1)^2} \sum_{w=v}^\II \bin wv 2^{-w}\\
&=\frac12 \sum_{v=b}^\II\frac{1}{(v+1)^2}2.
\end{align*}

In order to compute $T_1$, we now turn to the asymptotics of $\bin wv 2^{-w}$ for large  $w$. We obtain, by Stirling and setting $w=2v+\al$,
\bals
\bin wv 2^{-w}&\sim \frac{e^{-w}w^w\sqrt{2\pi w}}{e^{-(w-v)}(w-v)^{w-v}\sqrt{2\pi (w-v)}2^w e^{-v}v^v\sqrt{2\pi v}}\\
&=\frac{e^{-(2v+\al)}(2v+\al)^{2v+\al}\sqrt{2\pi (2v+\al)}}{e^{-(v+\al)}(v+\al)^{v+\al}\sqrt{2\pi (v+\al)}2^{2v+\al} e^{-v}v^v\sqrt{2\pi v}}\\
&\sim \frac{e^{-v}(2v)^{2v+\al}\lp 1+\frac{\al}{2v}\rp^{2v+\al} \sqrt{2}}{v^{v+\al}\lp 1+\frac{\al}{v}\rp^{v+\al}2^{2v+\al} e^{-v}v^v\sqrt{2\pi v}}\\
&\sim \frac{e^{\al+\frac{\al^2}{4v}}\sqrt{2}}{e^{\al+\frac{\al^2}{2v}}\sqrt{2\pi v}}\\
&\sim \frac{e^{-\frac{\al^2}{4v}}}{\sqrt{\pi v}}\\
&=2\frac{e^{-\frac{\al^2}{2\sig^2}}}{\sqrt{2\pi \sig}},
\end{align*}
with $\sig^2=2v$.
This is a Gaussian function, centered at $2v$ with variance $\sig^2=2v$. So, by Euler-Maclaurin, replacing sums by integrals, we obtain
\bit
\item if $b/2<v\leq b$,
\[\sum_{w=b}^\II \bin wv 2^{-w}\stackrel{b}{\sim} 2,\]
\item if $0\leq v<b/2$,
\[\sum_{w=b}^\II \bin wv 2^{-w}\mbox{ is exponentially negligible },\]
\item if $v \geq b$,
\[\sum_{w=b}^\II \bin wv 2^{-w}= 2  \mbox{ by (\ref{E5})},\]
but this will not be used in the sequel,
\eit
and finally
\[T_1+T_2\stackrel{b}{\sim} \frac12\lb   2\sum_{v=b/2} ^{b} \frac{1}{(v+1)^2}+   2\sum_{v=b} ^\II \frac{1}{(v+1)^2}  \rb\stackrel{b}{\sim} \frac2b.\]

This leads to
\beq
\E\lp \frac1R ;R>0\rp \stackrel{b}{\sim} \frac1L \lb \frac 2b -\frac1b           \rb=\frac{1}{Lb}.                 \label{E100}
\eeq
In the neighbourhood of $v=b/2$, only part of the Gaussian is integrated. But if we choose an interval $\De:= [b/2-b^{5/8},b/2+b^{5/8}]$,    
       ($ b^{5/8}\gg \sig$), this contributes to
\[\BO\lp \int_\De \frac{1}{v^2}dv\rp =\BO(b^{-5/8})=o(1/b).\]
Similarly, we derive (we omit the details)
\bals
\E\lp \frac{1}{R^2} ;R>0\rp &\stackrel{b}{\sim} \frac{3}{2Lb^2},\\
\V\lp \frac{1}{R^2} ;R>0\rp &\stackrel{b}{\sim} \frac{1}{b^2}\lb \frac{3}{2L}-\frac{1}{L^2}           \rb,\\
\E\lp \frac{1}{R^{1/2}} ;R>0\rp &\stackrel{b}{\sim}  2(\sqrt{2}-1)/(L\sqrt{b}).
\end{align*}
More generally,
\[\E\lp \frac{1}{R^{\al}} ;R>0\rp\stackrel{b}{\sim} \frac{1}{L\al b^\al}(2^\al-1),\al>0\]
\epr
Now we obtain, by (\ref{E1}) and Thm \ref{T1} the following Thm
\bth
The limiting distribution of $U/R$ for large $b$ is Gaussian.
\ethGL 
 Note that, by (\ref{E2}) and (\ref{E100}), we obtain
\bal
\E\lp \lp \frac{U}{R}\rp^2;R>0\rp &\stackrel{b}{\sim} p^2+\frac{pq}{Lb} ,\non \\ 
\V \lp \frac{U}{R};R>0\rp  &\stackrel{b}{\sim} \frac{pq}{Lb}.                                                   \label{E101}
\end{align}
This provides a confidence interval for $p$. With a confidence level of $5 \%$ for instance, we have
\[\lb \frac{U}{R}-2\sqrt{\frac{pq}{Lb}}\leq p \leq \frac{U}{R}+2\sqrt{\frac{pq}{Lb}}\rb,\]

and, as we can estimate $p$ by $\frac{U}{R}$, this leads to
\[\lb \frac{U}{R}-2\sqrt{\frac{\frac{U}{R}\lp 1-\frac{U}{R}\rp }{Lb}}\leq p \leq \frac{U}{R}+2\sqrt{\frac{\frac{U}{R}\lp 1-\frac{U}{R}\rp }{Lb}}\rb.\]

\subsection{Several Colors}       \label{S63}
If we are interested in the joint distribution of the statistic $U_1/R,\ldots  U_k/R$, which correspond to $k$ different colors among the present colors, we have an asymptotic conditional multinomial distribution. For instance, for $k=2$, this leads to
\[\bin{r}{u_1,u_2,r-u_1-u_2}p_1^{u_1}p_2^{u_2}(1-p_1-p_2)^{r-u_1-u_2},\]
with mean $rp_1,rp_2$.
So
\[\E\lp \frac{U_1}{R};R>0\rp=p_1,\E\lp \frac{U_2}{R};R>0\rp=p_2,\]
and we obtain similarly, conditioned on $R$
\bals
\E(U_1 U_2)&=R(R-1)p_1p_2,\\
\E\lp \lp \frac{U_1}{R}\rp \lp \frac{U_2}{R}\rp \rp &=p_1p_2-\frac{p_1p_2}{R},
\end{align*} 
and, unconditioning,
\[\E\lp \lp \frac{U_1}{R}\rp \lp \frac{U_2}{R}\rp;R>0\rp \stackrel{b}{\sim} p_1p_2-\frac{p_1p_2}{Lb},\]
or
\[Cov\lp \frac{U_1}{R},\frac{U_2}{R};R>0\rp \stackrel{b}{\sim}-\frac{p_1p_2}{Lb}.\]

\section{Multiplicities of colored keys}       \label{S7}
Counting the distinct number of keys is import for mining search engine data, see the paper by Kane et al \cite{watson}. It is important for search engines to correctly identify seasonal (colors) queries and make sure that their results are temporally appropriate. See the conference paper by Shokouhi, \cite{MSFT}. For example, queries with the key ``Wimbledon'' would take on a different color (season) throughout the year. In the winter we might return general information about the tennis tournament. In the spring perhaps logistics, travel and start date become more important. After the tournament starts the results of the matches become more relevant. During the championship match the local broadcast is the most relevant. One might be interested in estimating the size of these different seasons. How many tennis related queries do we expect to occur during the Wimbledon final?

Assume that the multiplicities of  color $C$ keys are given by iid random variables (RV), with distribution function $F_C$,  mean 
$\mu_C$, variance $\sigd_C$ (functions of $C$), all unknown. We want to estimate $\mu_C$ and $\sigd_C$. Of course, we can estimate 
$\mu_C$ by $\mut_C=N_C/n_C$ where $N_C$ is the total number of observed color $C$ keys and $n_C$ is the number of \emph{distinct}  color $C$ keys among the $n$ \emph{distinct} keys. $n_C$ is classically estimated by $2^D U$ (recall that  $U$ is the number of color $C$ keys among the $R$ distinct  keys in the cache). But the classical AS algorithm is not efficient enough to provide an estimate for 
$\sigd_C$. We proceed as follows: 
 to each color $C$  key $\K_i$, we attach  a counter giving its \emph{observed}  multiplicity $\mu_i$. 
  From Section \ref{S6} (see(\ref{E101})), we can estimate $p:=n_C/n$ by $\pt=(U/R;R>0)$. We have
\bals
\E(\pt;R>0)&\stackrel{b} {\sim} p,\\
\V(\pt;R>0)&\stackrel{b}{\sim} \frac{pq}{Lb}.
\end{align*} 
Also, we can estimate mean $\mu_C$ and  variance $\sigd_C$ by $\mut_C$ and $\sigt_C^2$ as given by (the observed multiplicities are 
extracted in the cache at  the end of AS)
\bals
\mut_C&:=\frac{V}{U},\B V:=\sum_1^U \mu_i,\\
\sigt^2_C&:=\frac{\sum_1^U (\mu_i-\mut_C)^2}{U}.
\end{align*} 
Next we estimate $n$ by $\nt =R2^D$  (see Sec. \ref{S2}).
We have, \emph {conditioned} on $U$,
\bth   \label {T71}
The moments of $\mut=V/U$   are given by     
\bals
\E(\mut)&=\mu,\\
\E(\mut^2)&=\mu^2+\sigd\E\lp \frac1U\rp,\\
\V(\mut)&=\sigd\E\lp \frac1U\rp.
\end{align*} 

\ethGL
\bpr
We only need
\[\E\lb \left.\frac{V^2}{U^2}\right|U \rb=\lb \left.\frac{U\sigd+U^2\mu^2}{U^2}\right|U \rb  \].
\epr
Now we estimate $n_C$ by $\nt_C=\nt\pt=2^D U$ . But if we have two independent RV, $X,Y$, with mean and variance respectively $m_X,m_Y,\sigd_X,\sigd_Y$, it is easy to see that
\bal
\E(XY)&=m_X m_Y,                                                 \label{E102}  \\
\V(XY)&=\sigd_X m^2_Y+\sigd_Y m^2_X + \sigd_X \sigd_Y.\non
\end{align}
 Here, our RV are not independent, but we can check that (\ref{E102}) is correct. The relation for the variances gives us a useful approximation. 
For instance
\bals
\E(\nt_C)&\stackrel{b}{\sim} np=n_C,
\mbox{ and, using (\ref{E21}),}\\
\V(\nt_C)&\sim \V(\nt)p^2+\V(\pt)n^2+\V(\nt)\V(\pt)\stackrel{b}{\sim}  \frac{n^2p(Lb-L+pL+1-p)}{L^2(b-1)b}\stackrel{b}{\sim}\frac{n^2p}{Lb} \mbox{ for large }b.
\end{align*}

It remains to estimate $\E\lp \frac1U\rp$ in order to complete $\E(\mut^2),\V(\mut)$. Using the binomial distribution $Bin(r,p)$ does not lead to a tractable expression. But, as $R$ is large whp, we can use the Gaussian approximation for $U$ as follows: conditioned on $R=r$, we have
\bals
\E\lp \frac1U\rp&\sim\int_1^r\frac{\exp\lp -\frac{(u-rp)^2}{2rpq}\rp}{\sqrt{2\pi rpq}u} du\\
&\sim \int_{-rp}^{rq} \frac{\exp\lp -\frac{v^2}{2rpq}\rp}{\sqrt{2\pi rpq}}\frac{1}{rp}\lp 1-\frac{v}{rp}+\frac{v^2}{r^2p^2}+\ldots\rp\\
&\sim \int_{-\II}^{\II} \frac{\exp\lp -\frac{v^2}{2rpq}\rp}{\sqrt{2\pi rpq}}\frac{1}{rp}\lp 1-\frac{v}{rp}+\frac{v^2}{r^2p^2}+\ldots\rp\\
&\sim \frac{1}{rp}\lp 1+\frac{q}{rp}\rp.
\end{align*}  
Unconditionning, this gives
\[\E\lp \frac1U;R>0\rp \stackrel{b}{\sim} \frac{1}{Lbp}+\frac{3q}{4p^2Lb^2}\stackrel{b}{\sim} \frac{1}{Lbp}\]
that we insert now into Thm \ref{T71}.
\section{The Black-Green  Sampling}       \label{S8}
This analysis was motivated by an oral question by P. Flajolet. 

In this section, we analyze  a case in some sense opposite to the one of Sec. \ref{S6}: here we do \emph{not} observe the color of each key. At first sight, all colors are black. Nevertheless, observing their multiplicities, we want to recover their colors at the end of the algorithm.

One difficulty encountered in clustering multidimensional data streams is in maintaining summaries of each cluster which are often space intensive. Methods such as {\em CSketch} have been developed which use a count-min sketch to store the frequencies of attribute-value combinations in each cluster, see the paper by C. Aggarwal \cite{csketch}  and \cite{AC2015}. This motivates the model we study below, where from a slight variant of AS, we recover estimates of the number of keys appearing with each frequency (hence their colors).

 We have two models: in Model I: only multiplicities can be observed,
in Model II AS is speeding up when we can observe \emph{one} extra color

Model I: only multiplicities can be observed

Assume that there are $n$ distinct keys, among which $n_i=n p_i(0<p_i<1)$ do have color $C_i,i=1\ldots k$. Assume also that each \emph{distinct} color $C_i$ key $\K_i$ appears $\mu_i$ times in the sample, all $\mu_i's$ being \emph{different}, but we can't observe the keys colors. So $N=\sum_1^k n_i\mu_i$. We want to estimate $n,n_i,\mu_i$, all unknown. (Here, we consider only the mean of our estimates).   We attach to each \emph{distinct} key 
$\K_\ell$ a counter $\nu_\ell$ giving the number of times (multiplicity)  this key appears in the sample. At the end of AS, we have in the cache $U_i$ distinct color $C_i$ keys, $i=1..k$, and each one will display the $same$ value for $\nu_i$, which is obviously equal to $\mu_i$. Hence the colors are now distinguished. 
This leads to the unbiased estimates $\nt_i=2^D U_i,\nt=2^D R,\pt_i=U_i/R$. (see Sec. \ref{S63}).

Model II: AS speeding up when we can observe \emph{one} extra color

Now we assume that we have an extra color  Green (G), with \emph{known} multiplicity  $\mu_G>\mu_i,i=1\ldots k$. ($\mu_i$ still unknown). We can improve the speed of AS as follows: at each depth $d,d=0..D$, each time a key obtains the value $\nu=\mu_G$, it is obviously G, and it is extracted from the cache. We have a vector counter $H$ such that, each time  a G key is extracted at depth  $d$ from the cache, $H[d]$ is increased by $1$. At the end of this new AS, the final number of process executions  is $D^*$, say, and we have the estimates
$\nt_i=2^{D^*}U_i,\mu_G\nt_G=N-\sum_1^k \mu_i\nt_i$, hence $\nt_G$, $\nt=\sum_1^k \nt_i+\nt_G,\pt_i=\nt_i/\nt$. A more precise estimate $\nt_G$ is obtained as follows: we use $\nt_G=\sum_0^{D^*}2^d H[d]$ (see the detailed explanation below).

Intuitively, $D^*<D$. To evaluate the difference $D-D^*$, we turn to an example.

 Assume that among the $n$ distinct keys,   $np$ ($0<p<1$) are Black (B), with   multiplicity $\mu_B$ and $nq,q:=1-p$ are Green (G), with \emph{known} multiplicity  $\mu_G> \mu_B$.  For instance, assume that each B key is unique and each G key   is present in triplicate. So we have a total of $N=np+3qn=n(3-2p)$ keys.  At the end of AS, each key with $\nu=1$ is obviously B.  As all $N$ keys are assumed to be distributed according to the uniform permutation distribution, we can consider the effect of each key on the cache as a Markov process: with probability $\frac{p}{3-2p}$, the key is B and it is inserted, with probability $\frac{3(1-p)}{3-2p}$, the key is G and three cases can occur: assume that the observed key appears in position $v,1\leq v\leq N$. Set $\ta:=v/N$. Then
 \bit
 \item With probability $\ta^2$, the key was the third  one among the three G keys with the same value, so it is deleted from the cache
 \item  With probability $2\ta(1-\ta)$, the key was the second   one, and it remains in the cache
 \item  With probability $(1-\ta)^2$, the key is the first   one, and it is inserted in the cache.
 \eit
 This can be seen as a Random walk on the cache. So the mean effect (on the cache size) of a G key at position $v$ is given by
 \[-\ta^2+0\times 2\ta(1-\ta)+1\times(1-\ta)^2=1-2\ta.\]
 Finally, the mean effect on the cache size of a  key at position $v$ is given by
 \[\pi(\ta)=\frac{p}{3-2p}+ \frac{3(1-p)}{3-2p}(1-2\ta)=\frac{3-6\ta-2p+6p\ta}{3-2p}.\]
 Consider now the process beginning  $(d=0)$. How many keys (in the mean) must be read in order to fill up the $b$ positions in the cache? This is given by $v_0$, where $b=V(0,v_0)$ and
 \[V(u_1,u_2)=\int_{u_1}^{u_2}\pi(\ta) dv=N \int_{u_1/N}^{u_2/N}\pi(\ta) d\ta= \frac{3(u_1^2-u_2^2)+3p(u_2^2-u_1^2)-3N(u_1-u_2)+2Np(u_1-u_2)}{N(3-2p)}.\]
 This leads to
 \[v_0= \frac{-3N+2Np+\lb N(-3+2p)(2Np-12bp-3N+12b)\rb^{1/2}}{6(p-1)}. \]
 An average of $b/2$ keys (starting with bit $1$) are killed for the next execution  $d=1$. But the mean  number of available keys still to be read is also divided by $2$. So the mean number of keys necessary to fill up the $b/2$ remaining positions in the cache is given by $v_1-v_0$, where $b/2=\frac12 V(v_0,v_1)$. This leads to
 \[v_1= \frac{-3N+2Np+\lb N(-3+2p)(2Np-24bp-3N+24b)\rb^{1/2}}{6(p-1)}. \]
 More generally, the mean number of keys necessary to fill up the $b/2$ remaining positions in the cache at depth $d$  is given by $v_d-v_{d-1}$, where $b/2=2^{-d}V(v_{d-1},v_d)$. This leads to
 \[v_d= \frac{-3N+2Np+\lb N(-3+2p)(2Np-2^d 12  bp-3N+2^d 12  b)\rb^{1/2}}{6(p-1)}, \]
 and finally, the mean total number $\E(D)$  is given by $\E(D)=\lceil D^* \rceil$, where $D^*$ is the solution of 
 \[N=\frac{-3N+2Np+\lb N(-3+2p)(2Np-2^{D^*} 12  bp-3N+2^{D^*} 12 b)\rb^{1/2}}{6(p-1)}.\]
 This gives
 \[D^*= \lg \lp  \frac{Np}{(3-2p)b}\rp=\lg N-\lg b +\lg p-\lg((3-2p))=\lg n-\lg b +\lg p.\]
 and
\[D-D^*=-\lg p>0.\]
This is the more positive the less $p$ is.  
 
 Note that, at the end, the number of B keys in the sample is estimated by
 \[2^{D^*} \times  \mbox{ number of  B keys in the cache},\]
obviously only B keys (with $\nu=1$) remain in the cache
 and the number of G keys in the sample is estimated by
 \[3.\sum_{d=0}^{D^*} 2^d H[d].\]
Indeed, imagine that we mark a key with a $*$ as soon as it is decided to be $G$ (because it is the third time we observe it). At depth $0$, $v\in [0,v_0)]$, all marked keys are counted in $H[0]$. At depth $1$, $v\in[v_0,v_1]$, all marked keys (starting with $0$)
are counted in $H[1]$, this corresponds in the mean, to $2H[1]$ $G$ keys, etc.
 Actually, the vector counter $H$ could be replaced by a single counter $H$ into which, at each depth $d$, we add the number of extracted G keys 
 $\times 2^d$.

\section{Conclusion}       \label{S10}
 Once again, the techniques using Gumbel-like distributions and Stein methodology proved to be quite efficient in the analysis of algorithms such as Adaptive Sampling.

\section*{Acknowledgements}
We would like to thank  J. Lumbroso  with whom we had many interesting discussions.
 We would also like to thank   the referee for careful reading and many useful suggestions
 that improved the paper.
 We would also like to thank Ben Karsin for the helpful discussions. 
 
 \bibliographystyle{plain} 

\appendix
\section{ Asymmetric Adaptive Sampling}       
For the sake of completeness, we analyze in this section the Asymmetric Adaptive Sampling. Assume that the hashing function gives asymmetric distributed bits.
Let $p$ denote the probability of bit $1$ ($q:=1-p$).
Now, the number of keys in the cache is  asymptotically Poisson with parameter $nq^d$ and the number of keys  in the twin bucket is 
asymptotically Poisson with parameter $npq^{d-1}=n\frac{p}{q} q^d$. So we set here
\bals
Q&:=1/q,\\
Z&:=\frac{R Q^D}{n},\\
\log &:=\log_Q,\\
\eta&:=d-\log n,\\
L &:=\ln Q,\\
\alt &:=\al/L,\\
\{x\}& :=\mbox{ fractional part of }x,\\
\chil &:=\frac{2l\pi \ii}{L}.
\end{align*} 
So the asymptotic distribution is now given by
\beq
p(r,d) \sim  f(r,\eta)= \exp(- e^{-L\eta} ) \frac{ e^{-Lr\eta}}{r!} 
\bigg[1-\exp(- e^{-L\eta} p/q) \sum_{k=0}^{b-r}  \frac{ e^{-Lk\eta}(p/q)^k}{k!} \bigg] ,                \label{E6}
\eeq
and
\[p(.,d):=\P(D=d)= \sum_{r=0}^b p(r,d).\]
This leads to

\[\FI(r,\al) =\int_{-\II}^\II e^{\al \eta} f(r,\eta)d\eta=
 \frac{\Gam(r-\alt)}{Lr!}
 -\sum_{k=0}^{b-r}\frac{\Gam( r+k-\alt )q^{r+k-\alt}(p/q)^k}{Lr!k! }.\]
In the sequel, we only provide the main related theorems. All detailed proofs can be found in this paper long version: 
\cite{LY17}.

\subsection{Moments of $D-\log n$}       \label{S70}
 We have

\bth
The asymptotic moments of $D-\log n$ are given by
 \bals
 \mt_{1,k}&=-\frac{\psi(k)}{L^2 k}+\sum_{i=0}^{b-k}\frac{(\psi(k+i)-L)q^k p^i\Gam(k+i)}{L^2\Gam(k+1)\Gam(i+1)},\B k>0,\\
 \mt_{1,0}&=\frac12+\frac{\gam}{L}+\sum_{i=1}^{b}\frac{(\psi(i)-L)p^i}{iL^2},\\
 \mt_{2,k}&=\frac{\psi(1,k)+\psi(k)^2}{L^3 k}+\sum_{i=0}^{b-k}-\frac{(-2\psi(k+i)L+L^2+\psi(1,k+i)+\psi(k+i)^2)q^kp^i\Gam(k+i)}{L^3\Gam(k+1)\Gam(i+1)},\B k>0,\\
 \mt_{1,0}&=\frac13+\frac{\gam}{L}+\frac{\pi^2}{6L^2}+\frac{\gam^2}{L^2}+\sum_{i=1}^{b}-\frac{(-2\psi(i)L+L^2+\psi(1,i)+\psi(i)^2)p^i}{iL^3},\\
  \wt_{1,k} &=\sum_{l\neq 0} \lb -\frac{\psi(k+\chil)\Gam(k+\chil)}{L^2\Gam(k+1)}+ \sum_{i=0}^{b-k}
  \frac{(\psi(k+i+\chil)-L)\Gam(k+i+\chil)q^{k+i}}{L^2\Gam(k+1)\Gam(i+1)} \rb e^{-2l\pi\ii \log n},\B k>0,\\
   \wt_{1,0}&=\sum_{l\neq 0}\lb  -\frac{\psi(\chil)\Gam(\chil)}{L^2}+ \sum_{i=0}^{b}
	\frac{(\psi(i+\chil)-L)\Gam(i+\chil)q^{i}}{L^2\Gam(i+1)} \rb e^{-2l\pi\ii \log n},\B k>0.
	\end{align*} 
	\ethGL

\subsection{Moments of $Z$}       \label{S71}
\bth
The non-periodic components of the moments of $Z$ are given by
\bals
 m_{1,k}&=
 1+\frac{(b-k)!}{L}\sum_{i=1}^{k-1}\Std{k}{i}\frac{q^{i-k}-1}{(k-i)(b-i)!},\\
 \V(Z)&\sim \frac{p}{(b-1)qL}.
 \end{align*}
 The periodic component is obtained as follows
 \bals
 w_{1,k}&= 
  \sum_{l\neq 0}  \frac1L\sum_{j=1}^{k-1}\Std{k}{j}
\lb(1-q^{j-k}\rb
\Gamma(j-k+\chil)\binom{b-k+\chil}{b-j}e^{-2l\pi\ii \log n}.
 \end{align*} 
\ethGL

\subsection{Distribution of $R$}       \label{S72}
\bth
The asymptotic distribution of $R$ is given by 
 \bals
 \P(R=r)&= p(r,.)\sim  \FI(r,0)
 = \frac1L\lb \frac1r- \sum_{u=r}^{b}\frac{(u-1)!}{r!(u-r)!}p^u(q/p)^r \rb,r\geq 1,
 \end{align*} 
\bals
 \E(R)&\sim 
 \frac1L\lb b-qb \rb =\frac{pb}{L},\\
 \E(R^2)&
 \sim\frac1L\lb b(b+1)/2-q^2\frac{b(b-1)}{2}-qb\rb.
 \end{align*} 
\ethGL

\subsection{Moments of $1/R,R>0 $  for large $b$  }   \label{S73}
\bth
The asymptotic moments of $1/R,R>0 $  for large $b$, with $R>0$ are given by 
\[\E\lp \frac1R ;R>0\rp \stackrel{b}{=} \frac{1}{Lb}\frac{p}{q}.\]
\ethGL

\end{document}